\documentclass[prc,superscriptaddress,unsortedaddress,twocolumn]{revtex4}
\usepackage{graphicx,epsf,bm,amsmath,color}
\usepackage{here,ulem}
\def\ba{\mbox{\boldmath $a$}}
\def\bb{\mbox{\boldmath $b$}}

\def\bk{\mbox{\boldmath $k$}}
\def\bp{\mbox{\boldmath $p$}}
\def\bq{\mbox{\boldmath $q$}}

\def\bfsigma{\mbox{\boldmath $\sigma$}}

\def\bftau{\mbox{\boldmath $\tau$}}
\def\sixj#1#2#3#4#5#6{\begin{Bmatrix}
                    #1 & #2 & #3 \\ #4 & #5 & #6 \end{Bmatrix} }
\def\ninj#1#2#3#4#5#6#7#8#9{\begin{Bmatrix}
                    #1 & #2 & #3 \\ #4 & #5 & #6 \\
                    #7 & #8 & #9  \end{Bmatrix}}
\begin{document}
\title{Partial-wave expansion of $\Lambda NN$ three-baryon interactions
in chiral effective field theory}
\author{M. Kohno}
\affiliation{Research Center for Nuclear Physics, Osaka University, Ibaraki 567-0047,
Japan}

\author{H. Kamada}
\affiliation{Department of Physics, Faculty of Engineering, Kyushu Institute of Technology,
Kitakyushu 804-8550, Japan}

\author{K. Miyagawa}
\affiliation{Research Center for Nuclear Physics, Osaka University, Ibaraki 567-0047,
Japan}

\begin{abstract}
An expression of partial wave expansion of three-baryon interactions in chiral effective
field theory is presented. The derivation follows the method by Hebeler \textit{et al.}
[Phys. Rev. C{\bf 91}, 044001 (2015)], but the final expression is more general.
That is, a systematic treatment of the higher-rank spin-momentum structure of
the interaction becomes possible. Using the derived formula, a $\Lambda$-deuteron folding
potential is evaluated. This information is valuable for inferring the possible contribution
of the $\Lambda NN$ three-baryon forces to the hypertriton as the basis of
further studies by sophisticated Faddeev calculations. A microscopic understanding
of $\Lambda NN$ three-baryon forces together with two-body $\Lambda N$
interactions is essential for the description of hypernuclei and neutron-star matter.
\end{abstract}

\maketitle
\section{Introduction}
Any description of two-body baryon-baryon interactions in which various
degrees of freedom are eliminated or frozen is effective. When the interactions
are applied in many-body systems, the appearance of three-body interactions is
inevitable as induced interactions. The important role of three-body forces (3BFs)
in nuclear physics has been observed in scattering and binding properties of
few-nucleon systems \cite{WGH98,SS02,PPW01} and also
in heavier nuclei and nuclear matter, in particular in connection with 
saturation properties \cite{PPW01,APR98,BB12}. The recent
development of the construction of baryon-baryon interactions in chiral effective
field theory (ChEFT) \cite{EHM09,ME11} provides a systematic way to introduce
three-body (and more-than-three-body) forces in a power-counting scheme and
therefore quantifies the role of 3BFs as opposed to simple phenomenological adjustment. 

The inclusions of 3BFs in a microscopic description of nuclei often need partial-wave
expansion in two Jacobi momenta. An efficient method was developed by
Hebeler \textit{et al.} \cite{Heb15} for the local 3BFs.
Here the local means that the interaction is a function
of the momentum transfer of each Jacobi momentum except for the cutoff
regularization function that does not depend on angle variables. In their method,
the original eight-dimensional angular integration, though five dimensional
because of the rotational invariance, was reduced essentially to two dimensional.

In this article, following the derivation in Ref. \cite{Heb15},
a different expression for the partial-wave expansion of 3BFs
is presented, which is more systematic for treating higher-rank coupling of
spin and momentum vectors.

Before discussing the partial-wave expansion, the basic structure of leading-order
3BFs in ChEFT is summarized in Sec. II. The expression of partial wave decomposition
of 3BFs in momentum space concerning the Jacobi momenta
is presented in Sec. III. As an application of the derived expression, a possible role of
the $\Lambda$NN 3BFs in the hypertriton is studied by calculating a $\Lambda$-deuteron
folding potential from $\Lambda NN$ 3BFs. A summary follows in Sec. IV.

\section{Structure of $\Lambda NN$ 3BFs in ChEFT}
\begin{figure}[b]
\centering
 \includegraphics[width=0.15\textwidth]{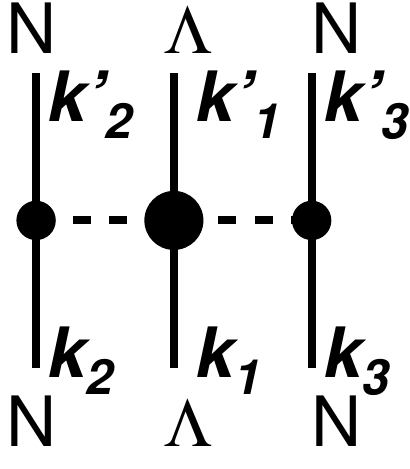}
\caption{Diagram of two-pion exchange $\Lambda NN$ 3BF. The small filled circle
denotes the $NN\pi$ vertex with the coupling constant $g_A/f_0^2$, and the large
filled circle denotes the $NN\pi\pi$ vertex specified by the coupling constants
$3b_0+b_D$ and $2b_2+3b_4$ in Eq. (\ref{eq:lnn}).}
\label{fig:lnn}
\end{figure}

Two-pion exchange $\Lambda NN$ 3BFs are considered as a concrete example,
which is relevant for studying hypertriton.
The structure of the $\Lambda NN$ force in the lowest-order,
namely next-to-next-to-leading order (NNLO), is particularly simple because the
$\pi\Lambda\Lambda$ vertex is not present. The contribution is only from
the diagram shown in Fig. \ref{fig:lnn}. The coordinate 1 is assigned to the
$\Lambda$ hyperon.

Following the expression by Petschauer \textit{et al.} \cite{Pet16}, the Born
amplitude of this diagram is written as
\begin{align}
 V_{TPE}^{\Lambda NN} = \frac{g_A^2}{3f_0^4} (\bftau_2\cdot\bftau_3)
\frac{(\bfsigma_3 \cdot \bq_{3d})(\bfsigma_2 \cdot \bq_{2d})}
{(\bq_{3d}^2+m_\pi^2)(\bq_{2d}^2+m_\pi^2)} \notag \\
 \times \{-(3b_0+b_D)m_\pi^2 +(2b_2+3b_4)\bq_{3d}\cdot\bq_{2d}\},
\label{eq:lnn}
\end{align}
where $\bq_{2d}$ ($\bq_{3d}$) is the difference between the final and initial momenta
at the nucleon line 2 (line 3): $\bq_{2d}=\bk_2'-\bk_2$ and $\bq_{3d}=\bk_3'-\bk_3$.
$g_A$ is the axial coupling constant, $f_0$ is the pion decay constant, $m_\pi$
is the pion mass, and $\bfsigma_i$ and $\bftau_i$ stand for the spin
and isospin operators, respectively, of nucleon $i$ (with $i=2,3$).
The coupling constants $b_0$, $b_D$, $b_2$, and $b_4$ inherit those in the
underlying Lagrangian. These coupling constants are to be determined in parametrizing
$\Lambda N$ interactions in the next-to-next-to-leading order. However, such an
attempt is not possible at present. In this paper, we use the estimation
by Petshauer \textit{et al.} \cite{Pet16}. 

\begin{figure}
\centering
 \includegraphics[width=0.4\textwidth]{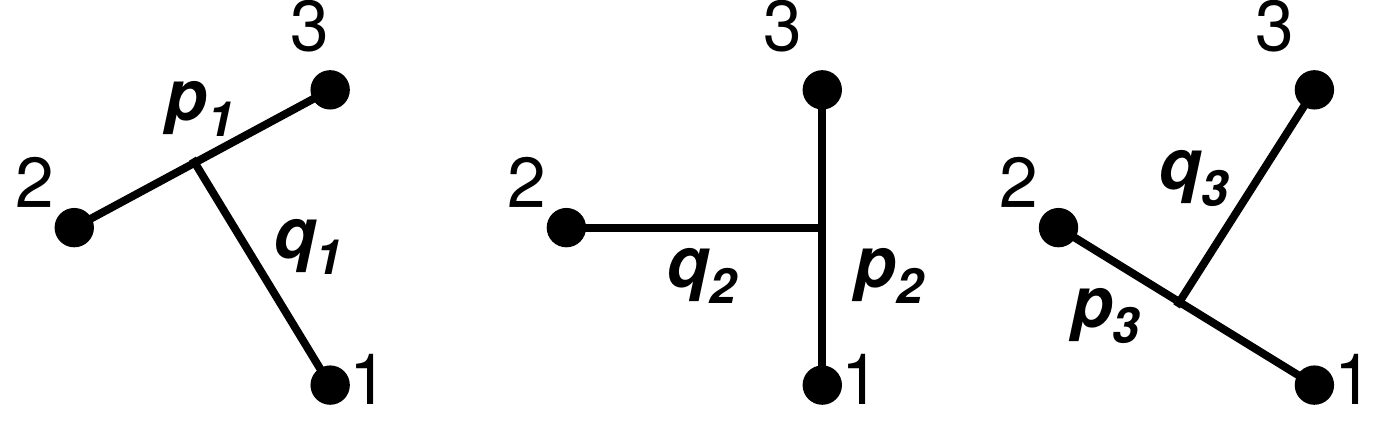}
\caption{Three types of the Jacobi momenta. The length of the vectors $\bp_i$
and $\bq_i$ does not correspond to the distance in the figure.}
\label{fig:jac}
\end{figure}
In the following, particle 1 is assigned to the $\Lambda$ hyperon, and the case of
the Jacobi momenta $(\bp_1,\bq_1)$ that is depicted by the leftmost
diagram of Fig. \ref{fig:jac} is considered. In the center-of-mass frame, $\bk_1=\bq_1$,
$\bk_2=\bp_1-r_{NN}\bq_1$, an $\bk_3=-\bp_1-r_{NN}\bq_1$ with $r_{NN}\equiv 1/2$.
Then, $\bq_{1d}=\bk_1'-\bk_1=\bq_1'-\bq_1$,
$\bq_{2d}=\bk_2'-\bk_2=\bp_1'-\bp_1-r_{NN}(\bq_1'-\bq_1)$, and
$\bq_{3d}=\bk_3'-\bk_3=-(\bp_1'-\bp_1)-r_{NN}(\bq_1'-\bq_1)$. $V_{TPE}^{\Lambda NN}$
is a function of $\bp\equiv \bp_1'-\bp_1$ and $\bq\equiv \bq_1'-\bq_1$ and can be
organized in the following tensor-product representation.
\begin{align}
 V_{TPE}^{\Lambda NN}(\bp,\bq) &= 4\pi (\bftau_2\cdot\bftau_3)\sum_{K=0,1,2}
 \sum_{\ell_a,\ell_b} V_{TPE}^{(K,\ell_a,\ell_b)}(p,q) \notag \\
 & \times [[Y_{\ell_a}(\hat{\bp})\times Y_{\ell_b}(\hat{\bq})]^K \times
 [\bfsigma_2\times \bfsigma_3]^K]_0^0,
\label{eq:tpe}
\end{align}
where the standard notation is employed for the tensor product:
\begin{align}
 [Y_{\ell_a}(\hat{\ba})\times Y_{\ell_b}(\hat{\bb})]_{m_c}^{\ell_c}=\sum_{m_a m_b}
 (\ell_a m_a \ell_b m_b|\ell_c m_c)\notag \\
 \times Y_{\ell_a m_a}(\hat{\ba}) Y_{\ell_b m_b}(\hat{\bb}).
\end{align}
The explicit expressions of $V_{TPE}^{(K,\ell_a,\ell_b)}(p,q)$ for the Jacobi
momenta $(\bp_1,\bq_1)$ are given in Appendix A. It is straightforward
to obtain a similar representation for the other two sets of the Jacobi
momenta, $(\bp_2,\bq_2)$ and $(\bp_3,\bq_3)$ in Fig. 2, for the 3BFs
of Eq. (\ref{eq:lnn}). Because the mass of the $\Lambda$ hyperon differs from that
of the nucleon, $\bk_i$ and $\bk_i'$ ($i=1,2,3$) are not treated cyclically
and the functional form of  $V_{TPE}^{K,\ell_a,\ell_b}(p,q)$ is different from
those given in Appendix A. It is noted that the third sigma operator $\bfsigma_3$
may appear in a general 3BF. In that case, the rank of $K=3$ can appear.

As for the cutoff, the following regulator function is introduced for the initial
and final Jacobi momenta, $(\ba, \bb)=(\bp_1,\bq_1)$ or $(\bp_1',\bq_1')$,
with the scale of $\Lambda=550$ MeV in present calculations:
\begin{equation}
 f_\Lambda(\ba,\bb)= \exp \{-(\ba^2 +\frac{3}{4}\bb^2)^2/\Lambda^4\}.
\label{Eq:regu}
\end{equation}
Because this function does not depend on the angles, it does not affect the
calculation of the partial-wave expansion, which is discussed in the next section.

\section{Partial-wave expansion}
A 3BF is in general a function of the initial and final Jacobi momenta,
$V_{3BF}(\bp_1', \bq_1',\bp_1,\bq_1)$ in the case of the leftmost diagram of Fig. 2
with supressing the spin and isospin indices.
The partial-wave expansion requires integrals of the product of four spherical
harmonics and $V_{3BF}(\bp_1', \bq_1',\bp_1,\bq_1)$ over the solid angles
related to Jacobi momenta $\bp_1'$, $\bq_1'$, $\bp_1$, and $\bq_1$ for
$V_{3BF}(\bp_1', \bq_1',\bp_1,\bq_1)$:
\begin{align}
 \frac{1}{\{(2\pi)^{3/2}\}^4} \int\cdots\int d\hat{\bp_1'}
d\hat{\bq_1'} d\hat{\bp_1} d\hat{\bq_1} Y_{\ell_p' m_p'}^*(\hat{\bp_1'}) \notag \\
 \times Y_{\ell_q' m_q'}^*(\hat{\bq_1'}) Y_{\ell_p m_p}(\hat{\bp_1}) Y_{\ell_q m_q}(\hat{\bq_1})
 V_{3BF}(\bp_1', \bq_1',\bp_1,\bq_1).
\end{align}
To make the expression compact, the following notation is used in the
subsequent derivation, in which the angular momenta $\ell_p$ and $\ell_q$ are coupled
to $L$, and $\ell_p'$ and $\ell_q'$ to $L'$, respectively. That is, supplementing the summation
\begin{equation}
 \sum_{m_{\ell_p'}m_{\ell_q'}} \sum_{m_{\ell_p}m_{\ell_q}} (\ell_p' m_{\ell_p'} \ell_q' m_{\ell_q'}|L' M_L')
 (\ell_p m_{\ell_p} \ell_q m_{\ell_q}|L M_L),
\end{equation}
the angular integration on a 3BF $V_{3BF}$ is expressed as
\begin{align}
 &\frac{1}{(2\pi)^6}
 \langle [Y_{\ell_p'}(\hat{\bp_1'})\times Y_{\ell_q'}(\hat{\bq_1'})]_{M_L'}^{L'}| V_{3BF} \notag \\
 &\times |[Y_{\ell_p}(\hat{\bp_1})\times Y_{\ell_q}(\hat{\bq_1})]_{M_L}^{L} \rangle,
 \label{eq:pwe}
\end{align}
where the left- and right-angle brackets represent $d\hat{\bp_1'}
d\hat{\bq_1'} d\hat{\bp_1} d\hat{\bq_1}$ integration.

The angular momentum projection in momentum space postulates a complete
plane-wave basis \cite{GL96},
\begin{equation}
 \langle \bp' |p\ell_p m_p \rangle =\frac{\delta(p'-p)}{p'p} Y_{\ell_p m_p}(\hat{\bp'}).
\end{equation}
The three-body basis states in jj coupling for the total three-body 
angular momentum $J$ are constructed as
\begin{align}
 |pq\alpha\rangle \equiv & |pq(\ell_p s_p)j_p (\ell_q 1/2) j_q (j_p j_q)JM\rangle \notag \\
   =& \sum_{m_{j_p} m_{j_q}} \sum_{m_p m_{s_p}} \sum_{m_q m_{s_q}} (j_p m_{j_p} j_q m_{j_q}|JM)
 \notag \\
  & \times (\ell_p m_p s_p m_{s_p}|j_p m_{j_p}) (\ell_q m_q s_q m_{s_q}|j_p m_{j_p}) \notag \\
  & \times |p\ell_p m_p\rangle \chi_p^{s_p,m_{s_p}} |q\ell_q m_q\rangle \chi_q^{s_q,m_{s_q}},
\end{align}
where $\chi_p^{s_p,m_{s_p}}$ and $\chi_q^{s_q,m_{s_q}}$ represent spin states of the $\bp$
and $\bq$ degrees of freedom, rspectively. The isospin state can be treated separately.
The basis states satisfy the orthonormality condition:
\begin{equation}
  \langle p'q'\alpha' |pq\alpha\rangle= \frac{\delta(p'-p)}{p'p}
  \frac{\delta(q'-q)}{q'q}\delta_{\alpha'\alpha}.
\end{equation}

For the case of a local 3BF
$V_{3BF}(\bp_1', \bq_1',\bp_1,\bq_1)= V_{3BF}(\bp_1'-\bp_1, \bq_1'-\bq_1)$,
the subtle manipulation \cite{Heb15} of adding a radial part to the angle integration
while keeping the absolute value by a delta function is helpful,
\begin{align}
 d\hat{\bp_1'} \rightarrow d\bp_1'' \frac{\delta(p_1''-p_1')}{{p_1'}^2} \hspace{0.5em}
 \mbox{and}\hspace{0.5em}d\hat{\bq_1'} \rightarrow d\bq_1''\frac{\delta(q_1''-q_1')}{{q_1'}^2}.
\end{align}
Changing the variables $\bp_1''$ and $\bq_1''$ to $\bp$ and $\bq$ by the shifts
of $\bp_1''=\bp+\bp_1$ and $\bq_1''=\bq+\bq_1$, Eq. (\ref{eq:pwe}) is modified as
\begin{align}
  & \frac{1}{(2\pi)^6}
 \langle [Y_{\ell_p'}(\hat{\bp_1'})\times Y_{\ell_q'}(\hat{\bq_1'})]_{M_L'}^{L'}| V_{3BF} \notag \\
 &\times |[Y_{\ell_p}(\hat{\bp_1})\times Y_{\ell_q}(\hat{\bq_1})]_{M_L}^{L} \rangle \notag \\
 =& \frac{1}{({2\pi})^6} \int_0^\infty p^2dp  \int_0^\infty q^2 dq \notag \\
 &\hspace{2em} \times \langle [Y_{\ell_p'}(\widehat{\bp_1+\bp})\times
  Y_{\ell_q'}(\widehat{\bq_1+\bq})]_{M_L'}^{L'}| \notag \\
 &\hspace{2em} \times
 \frac{\delta(|\bp-\bp_1|-p_1')}{{p_1'}^2} \frac{\delta(|\bq-\bq_1|-q_1')}{{q_1'}^2} V_{3BF}
 \notag \\
 &\hspace{2em}\times |[Y_{\ell_p}(\hat{\bp_1})\times Y_{\ell_q}(\hat{\bq_1})]_{M_L}^{L}\rangle.
 \label{eq:pwef}
\end{align}

The delta function can be written as follows by using a Legendre polynomial
of the first kind $P_k$:

\begin{align}
 & \delta(|\bp-\bp_1|-p_1')= \frac{p_1'}{pp_1}
  \delta\left(\cos\widehat{\bp_1\bp}-\frac{p_1'^2-p_1^2-p^2}{2p_1p}\right)\notag \\
  =& 2\pi \frac{p_1'}{pp_1} \sum_{k'} P_{k'}(c_p) (-1)^k \sqrt{\hat{k'}}
 [Y_{k'}(\hat{\bp_1})\times Y_{k'}(\hat{\bp})]_0^0,
\end{align}
\begin{align}
 & \delta(|\bq-\bq_1|-q_1')= \frac{q_1'}{qq_1}
   \delta\left(\cos\widehat{\bq_1\bq}-\frac{q_1'^2-q_1^2-q^2}{2q_1q}\right)\notag \\
 =&2\pi \frac{q_1'}{qq_1} \sum_{k} P_{k}(c_q) (-1)^k \sqrt{\hat{k}}
 [Y_{k}(\hat{\bq_1})\times Y_{k}(\hat{\bq})]_0^0,
\end{align}
where $\hat{k}\equiv 2k+1$, $c_p \equiv \frac{p_1'^2-p_1^2-p^2}{2p_1p}$, and
$c_q\equiv \frac{q_1'^2-q_1^2-q^2}{2q_1q}$.
These $\delta$ functions restrict the $p$ and $q$ integrations as
\begin{align}
 p_{min}\equiv |p_1'-p_1| \le p \le p_{max}\equiv p_1'+p_1, \\
 q_{min}\equiv |q_1'-q_1| \le q \le q_{max}\equiv q_1'+q_1.
\end{align}
The spherical-harmonic functions $Y_{\ell_p'}(\widehat{\bp_1+\bp})$ and
$Y_{\ell_q'}(\widehat{\bq_1+\bq})$ are also expanded as follows:
\begin{align}
 Y_{\ell_p' m_p'}(\widehat{\bp_1+\bp})=& \sum_{j_p+j_p'=\ell_p'}
 \sqrt{\frac{4\pi (\hat{\ell_p'})!}{(\hat{j_p})!(\hat{j_p'})!}}\frac{p_1^{j_p} p^{j_p'}}{{p_{1}'}^{\ell_p'}}
 \notag\\
 & \times [Y_{j_p}(\hat{\bp_1})\times Y_{j_p'}(\hat{\bp})]_{m_p'}^{\ell_p'}, \\
 Y_{\ell_q' m_q'}(\widehat{\bq_1+\bq})=& \sum_{j_q+j_q'=\ell_q'}
 \sqrt{\frac{4\pi (\hat{\ell_q'})!}{(\hat{j_q})!(\hat{j_q'})!}}\frac{q_1^{j_q} q^{j_q'}}{{q_{1}'}^{\ell_q'}}
 \notag\\
 & \times [Y_{j_q}(\hat{\bq_1}) \times Y_{j_q'}(\hat{\bq})]_{m_q'}^{\ell_q'}.
\end{align}

\begin{widetext}
A straightforward evaluation of the recoupling of these decoupled spherical harmonics
finally gives
\begin{align}
 & \frac{1}{(2\pi)^6}\langle [Y_{\ell_p'}(\hat{\bp_1'})\times Y_{\ell_q'}(\hat{\bq_1'})]_{M_L'}^{L'}|
  V_{TPE}^{\Lambda NN}
  |[Y_{\ell_p}(\hat{\bp_1})\times Y_{\ell_q}(\hat{\bq_1})]_{M_L}^{L} \rangle \notag \\
 = & \frac{1}{({2\pi})^6} (-1)^{\ell_p' +\ell_q'-L'} \sum_{JM} (-1)^{M_L'}
  (L'-M_L' LM_L|JM) \frac{1}{p_{1}'q_{1}'{p_1}{q_1}} \int_{p_{min}}^{p_{max}} p dp
 \int_{q_{min}}^{q_{max}} q dq \notag \\
 & \times (2\pi)^2 \sum_{k'k}  \hat{k'}\hat{k} P_{k'}(c_p) P_{k}(c_q) \sum_{j_p+j_p'=\ell_p'}
 \sqrt{\frac{(\hat{\ell_p'})!}{(\hat{j_p})!(\hat{j_p'})!}}\frac{p_1^{j_p} p^{j_p'}}{{p_{1}'}^{\ell_p'}}
 \sum_{j_q+j_q'=\ell_q'} (-1)^{j_p+j_q} \sqrt{\frac{ (\hat{\ell_q'})!}{(\hat{j_q})!(\hat{j_q'})!}}
 \frac{q_1^{j_q} q^{j_q'}}{{q_{1}'}^{\ell_q'}} \sqrt{\hat{\ell_p'} \hat{\ell_q'}\hat{L}\hat{L'}} \notag \\
 & \times 
 \frac{1}{4\pi} \sum_{L_p L_q}\sqrt{\hat{j_p'}\hat{j_q'}\hat{j_p}\hat{j_q}}
  (k'0 j_p'0|L_p0) (k0j_q'0|L_q0) (k'0 j_p0|\ell_p0) (k0j_q0|\ell_q0) \notag \\
 & \times
  \sixj{\ell_q}{\ell_q'}{L_q}{j_q'}{k}{j_q} \sixj{\ell_p}{\ell_p'}{L_p}{j_p'}{k'}{j_p}
 \ninj{\ell_p'}{\ell_q'}{L'}{\ell_p}{\ell_q}{L}{L_p}{L_q}{J} \int d\hat{\bp} d\hat{\bq}\;
 [Y_{L_p}(\hat{\bp})\times Y_{L_q}(\hat{\bq})]_M^{J} V_{TPE}^{\Lambda NN}(\bp,\bq)
 \label{eq:final}
\end{align}
\end{widetext}
Observing $P_\ell(\cos \widehat{\bp\bq})=(-1)^\ell \frac{4\pi}{\sqrt{\hat{\ell}}}
 [Y_\ell(\hat{\bp})\times Y_\ell(\hat{\bq})]_0^0$, the above expression with $J=0$
corresponds to Eq. (6) of Ref. \cite{Heb15}.
It is verified numerically that Eq. (\ref{eq:final}) with $J=0$ delivers
the same results as Eq. (6) of Ref. \cite{Heb15}.
The evaluation of $\hat{I}_{L'L}V_{3BF}$ with $J  > 0$ is straightforward
and transparent.

The angle integration in Eq. (\ref{eq:final}) for 3BFs in the form
of Eq. (\ref{eq:tpe}) needs

\begin{align}
   \int d\hat{\bp} d\hat{\bq} \; [Y_{L_p}(\hat{\bp})\times Y_{L_q}(\hat{\bq})]_M^{J}
 [Y_{\ell_a}(\hat{\bp})\times Y_{\ell_b}(\hat{\bq})]_\mu^K \notag \\
 = (-1)^{K+\mu} \delta_{JK} \delta_{M,-\mu}\delta_{L_p \ell_a}
 \delta_{L_q \ell_b}(-1)^{\ell_a+\ell_b}.
\end{align}
This means that the angular momentum $J$ in Eq. (\ref{eq:final}) corresponds to
the rank $K$ of the tensor-product structure of the angular-momentum coupling
in 3BFs. 

For actual calculations of the matrix element of 3BFs in various situations, it is
convenient to define a reduced matrix element in a form similar to the
Wigner-Eckart theorem,
\begin{widetext}
\begin{align}
 & \frac{1}{(2\pi)^6}\langle [Y_{\ell_p'}(\hat{\bp_1'})\times Y_{\ell_q'}(\hat{\bq_1'})]_{M_L'}^{L'}|
 V_{TPE}^{(K,\ell_a,\ell_b)}(p,q) [Y_{\ell_a}(\hat{\bp})\times Y_{\ell_b}(\hat{\bq})]_\mu^K
 |[Y_{\ell_p}(\hat{\bp_1})\times Y_{\ell_q}(\hat{\bq_1})]_{M_L}^{L} \rangle \notag \\
 =&(LM_L K\mu|L'M_L') \langle [Y_{\ell_p'}(\hat{\bp_1'})\times Y_{\ell_q'}(\hat{\bq_1'})]^{L'}||
V_{TPE}^{(K,\ell_a,\ell_b)}(p,q) [Y_{\ell_a}(\hat{\bp})\times Y_{\ell_b}(\hat{\bq})]^K
  ||[Y_{\ell_p}(\hat{\bp_1})\times Y_{\ell_q}(\hat{\bq_1})]^{L} \rangle_{pwe}.
\end{align}
From Eq. (\ref{eq:final}), the reduced matrix element is found as
\begin{align}
 & \langle [Y_{\ell_p'}(\hat{\bp_1'})\times Y_{\ell_q'}(\hat{\bq_1'})]^{L'}||V_{TPE}^{(K,\ell_a,\ell_b)}(p,q)
 [Y_{\ell_a}(\hat{\bp})\times Y_{\ell_b}(\hat{\bq})]^K
  ||[Y_{\ell_p}(\hat{\bp_1})\times Y_{\ell_q}(\hat{\bq_1})]^{L} \rangle_{pwe} \notag \\
 =& \frac{1}{4\pi} \frac{1}{(2\pi)^4}\sqrt{\frac{\hat{K}}{\hat{L'}}} (-1)^{\ell_p+\ell_q}
 \frac{1}{p_{1}'q_{1}'{p_1}{q_1}}
 \sqrt{\hat{\ell_p'}\hat{\ell_q'}\hat{\ell_p}\hat{\ell_q}\hat{L}\hat{L'}\hat{\ell_a}\hat{\ell_b}}
 \int_{p_{min}}^{p_{max}} p dp \int_{q_{min}}^{q_{max}} q dq
 \sum_{k'k} \hat{k'}\hat{k}P_{k'}(c_p)P_{k}(c_q)\notag \\
 & \times  \sum_{j_p+j_p'=\ell_p'}
 \sqrt{\frac{(\hat{\ell_p'})!}{(\hat{j_p})!(\hat{j_p'})!}}\frac{p_1^{j_p} p^{j_p'}}{{p_{1}'}^{\ell_p'}}
 \sum_{j_q+j_q'=\ell_q'} (-1)^{j_p+j_q}\sqrt{\frac{ (\hat{\ell_q'})!}{(\hat{j_q})!(\hat{j_q'})!}}
 \frac{q_1^{j_q} q^{j_q'}}{{q_{1}'}^{\ell_q'}}  (k'0 \ell_a0|j_p'0) (k0\ell_b0|j_q'0) \notag \\
 & \times (k'0 \ell_p0|j_p0) (k0\ell_q0|j_q0) \sixj{\ell_q}{\ell_q'}{\ell_b}{j_q'}{k}{j_q}
 \sixj{\ell_p}{\ell_p'}{\ell_a}{j_p'}{k'}{j_p} \ninj{\ell_p'}{\ell_q'}{L'}{\ell_p}{\ell_q}{L}{\ell_a}{\ell_b}{K}
 V_{TPE}^{(K,\ell_a,\ell_b)}(p,q)
 \label{eq:redm}.
\end{align}
\end{widetext}
It should be remembered that this expression is valid for a 3BF depending only
on $\bp=\bp_1'-\bp_1$ and $\bq_1'-\bq_1$. Introduction of the regularization
functions given by Eq. (\ref{Eq:regu}) does not affect the angular integrations. 

\section{$\Lambda$-deuteron folding potential}
The construction of hyperon-nucleon interactions in the strangeness $S=-1$
sector has a difficulty in lacking sufficient scattering data.
The fact that there is no two-body $\Lambda N$ bound state enhances the difficulty.
The hypertriton is, therefore, an important hyper-nuclear system \cite{MK95,HMN20}
for adjusting the interaction, where the tuning of $\Lambda N$ interactions in the
spin-singlet and -triplet states together with the strength of $\Lambda N$-$\Sigma N$
coupling can be carried out. However, the possible role of the $\Lambda NN$
3BFs has been investigated little. Expecting the settlement of the current
controversy over the shallow binding energy \cite{STAR00}, it is important to
describe the hypertriton including $\Lambda NN$ 3BFs.
Before considering Faddeev calculations for
the hypertriton including $\Lambda NN$ 3BFs, it is instructive to
calculate the $\Lambda$-deuteron folding potential due to the $\Lambda NN$
interactions, applying the expression derived in the previous section,
to obtain some idea about the contribution of the 3BFs.

The folding potential is evaluated by the following integration:
\begin{align}
 U_{\Lambda-d}^{J_t}(q_1',q_1)=& \iint p_1'^2 dp_1' p_1^2 dp_1
 \langle [\Psi_d(\bp_1'),(\ell_{\Lambda}' 1/2)j_\Lambda]J_t|
  \notag \\
  & \times V_{TPE}^{\Lambda NN}| [\Psi_d(\bp_1),(\ell_{\Lambda} 1/2)j_\Lambda]J_t
   \rangle, \label{eq:lpd} \\
 \Psi_d(\bp_1)=& \sum_{\ell_d=0,2} \frac{1}{p_1}\phi_{\ell_d}(p_1) [Y_{\ell_d}(\hat{\bp_1})
 \times \chi_d^{1}]_{m}^1.
\label{eq:sd}
\end{align}
The above expression is in an abbreviated notation. A more detailed calculational
procedure is given in Appendix B.

\begin{figure}[b]
\centering
 \includegraphics[width=0.4\textwidth,clip]{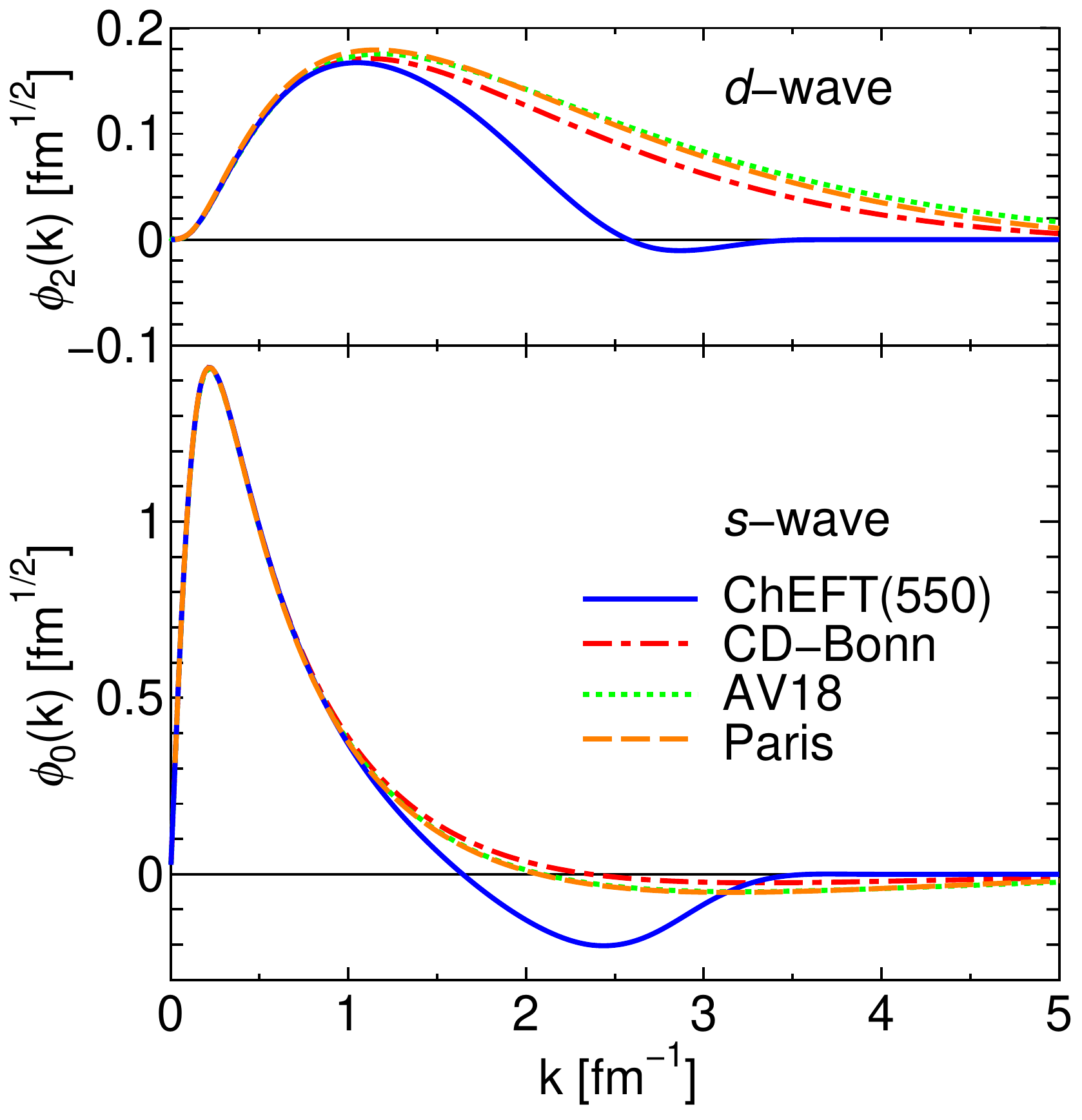}
 \caption{Deuteron $s$- and $d$-wave functions in momentum space described by various
$NN$ interactions: ChEFT \cite{Epe05}, AV18 \cite{AV18}, CD-Bonn \cite{CDB},
and Paris \cite{PARIS}. The sign of the $d$-wave function is reversed.
The normalization of these wave functions is
$\int_0^\infty dk\:(\phi_{0}^2(k)+\phi_{2}^2(k))=1$.}
\label{fig:dwf}
\end{figure}

Deuteron wave functions, the $s$ and $d$ components, are those of the N$^3$LO
ChEFT interactions with the cutoff of 550 MeV \cite{Epe05}. This scale of the wave
functions and the 3BFs may not be soft enough to permit a perturbative treatment.
Nevertheless, without strong short-range singularities, the resulting folding potential helps
intuitively infer the $\Lambda$-deuteron interaction and therefore
the possible role of the $\Lambda NN$ 3BFs to the hypertriton. It is noted that
it is not appropriate to employ deuteron wave functions of other $NN$ interactions
having strong short-range singularities together with the ChEFT 3BFs.  For comparison,
deuteron wave functions in momentum space are shown in Fig. \ref{fig:dwf}
in which those of other $NN$ interactions, i.e., AV18 \cite{AV18}, CD-Bonn \cite{CDB},
and Paris \cite{PARIS}, are compared.

\begin{figure}[b]
\centering
 \includegraphics[width=0.4\textwidth,clip]{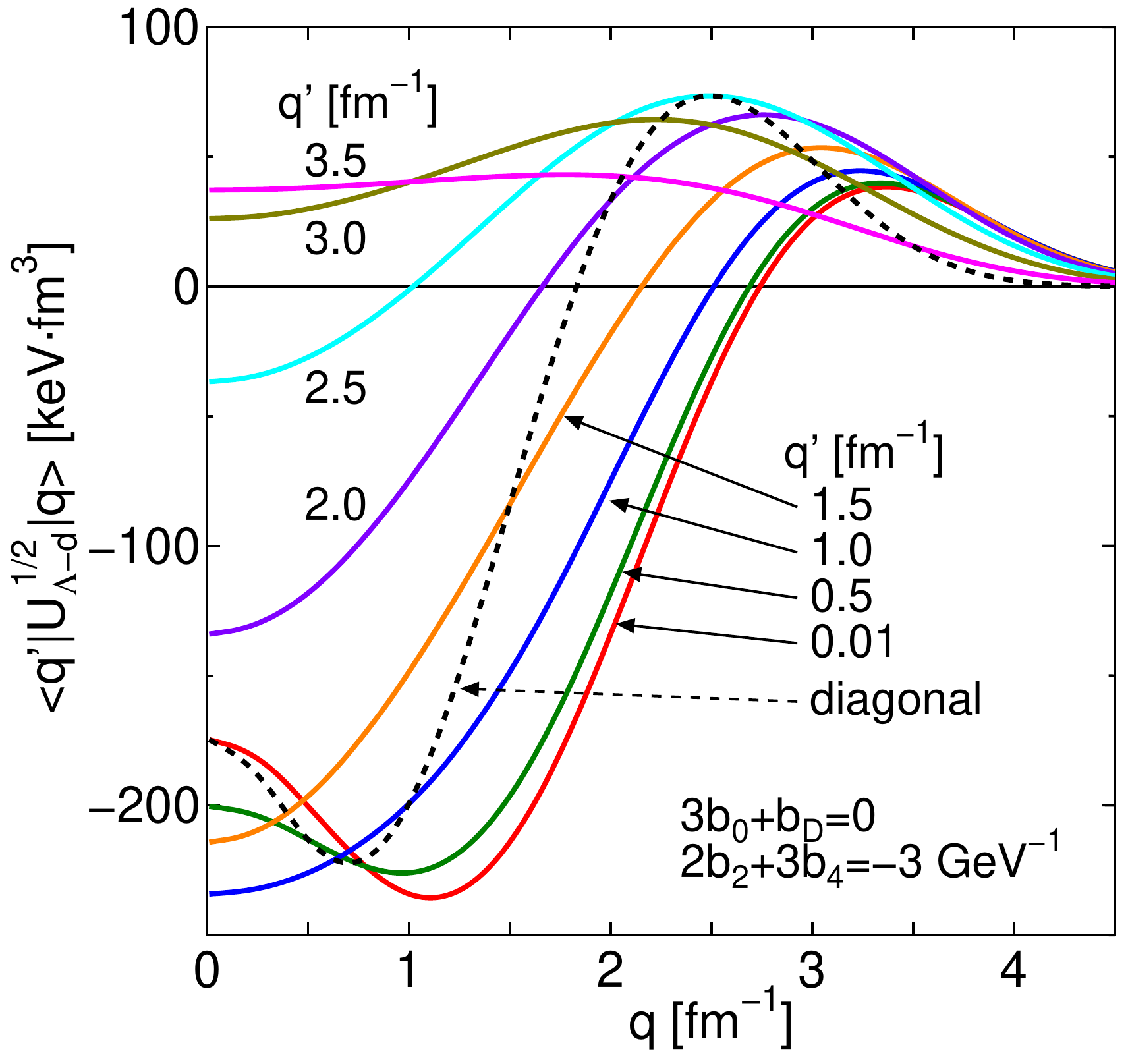}
 \includegraphics[width=0.4\textwidth,clip]{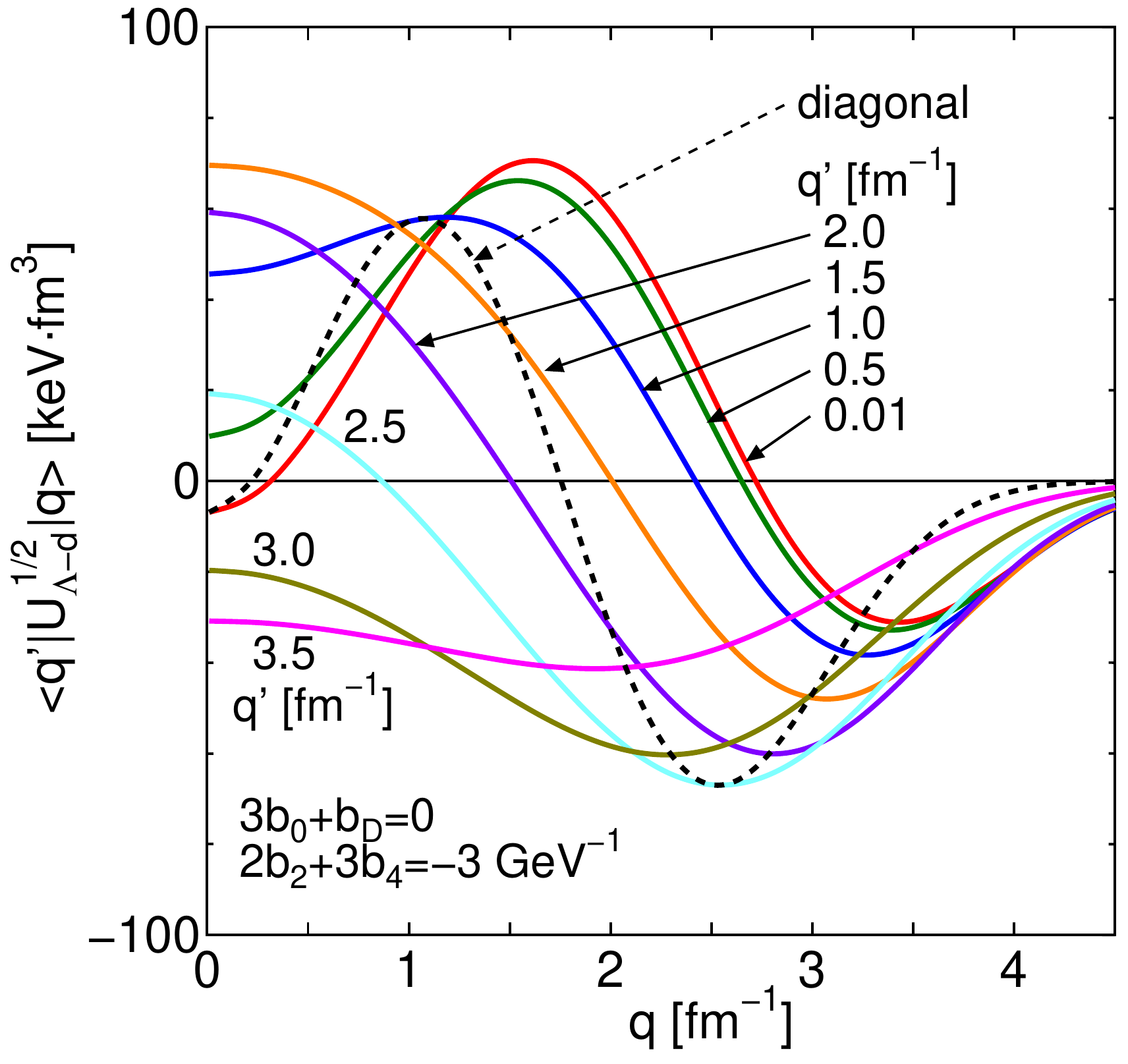}
 \caption{$\Lambda$-deuteron folding potential with $\ell_\Lambda=0$ from two-pion
exchange $\Lambda NN$ interactions. The upper panel shows the contribution of
the deuteron $s$-state pair. The lower panel shows the contribution of
from the remaining pairs: $sd$, $ds$, and $dd$.}
\label{fig:ldeut}
\end{figure}

Calculated $s$-wave [$\ell_{\Lambda}'=\ell_{\Lambda}=0$ in Eq. (\ref{eq:lpd})]
$\Lambda$-deuteron folding potential with $J_t=1/2$ from the leading-order
$\Lambda NN$ interactions is presented in Fig. \ref{fig:ldeut}. 
The upper panel shows the contribution from the $s$-wave component of the deuteron
wave function, in which $V_{TPE}^{(K=0,\ell_a,\ell_b)}$ participates. The lower panel presents
the sum of the remaining contributions from the $s$-$d$, $d$-$s$, and $d$-$d$ pairs of
the deuteron wave function. $V_{TPE}^{(K=2,\ell_a,\ell_b)}$ contributes in the $s$-$d$
and $d$-$s$ pairs. Both $V_{TPE}^{(K=0,\ell_a,\ell_b)}$ and $V_{TPE}^{(K=2,\ell_a,\ell_b)}$
contribute in the case of the $d$-$d$ pair.
The potential with $J_t=3/2$ is identical to that with $J_t=1/2$ . 

The employed parameters are taken from the estimation by Petshauer \textit{et al.}
\cite{Pet16}; that is, $(3b_0+b_D)=0$ and $(2b_2 +3b_4)=3.0\times 10^{-3}$
MeV$^{-1}$.
In principle, parameters of the two-pion exchange 3BFs should be determined
in fitting the two-body $\Lambda N$ interactions. However, the present
experimental situation of the strangeness $S=-1$ sector does not allow such
an investigation. 

The calculated potential is weakly attractive, with a depth below
about 200 keV. The experimental separation energy of the hypertriton
is very small, $130\pm 50$ keV, though the actual value is still controversial.
Therefore, its wave function extends and the $\Lambda NN$ 3BF contribution
can be considered hindered. Still, the similar magnitude of the $\Lambda NN$ 3BF
effect shown in Fig. \ref{fig:ldeut} as the separation energy indicates that the
effect may not be negligible in the hypertriton. Thus, it suggests that further
study of the hypertriton in Faddeev formalism with incorporation of the 3BFs
is necessary. The repulsive bump structure seen beyond $q \simeq
2.5$ fm implies that the scale of 550 MeV employed for the ChEFT description
still has remnants of shorter-range singularities, which should be treated in a
Faddeev framework.   

\bigskip
\section{Summary}
We have presented an expression of partial-wave decomposition of 3BFs concerning
the relevant Jacobi momenta. The derivation essentially follows that of Hebeler
\textit{et al.} \cite{Heb15}, but the final formula differs in that it can systematically treat
the higher-rank spin and angular-momentum tensor-product structure of 3BFs.
Although the consideration is intended specifically for $\Lambda NN$ 3BFs and one set
of the Jacobi momenta, the formula is general, as far as 3BFs are a function
of the momentum transfer in each Jacobi momentum. Even if a regularization
function is angle-dependent, the 3BFs can be expressed in the form of Eq. (\ref{eq:tpe})
and then the result of Eq. (\ref{eq:redm}) is applicable.

As an application of the derived expression, the $\Lambda$-deuteron folding potential
from NNLO $\Lambda NN$ 3BFs is evaluated. At the present stage, the construction
of baryon-baryon interactions in the strangeness $S=-1$ sector in ChEFT is practiced
up to the NLO level \cite{NLO13,HMN20}. Even at this low order,
it is difficult to unambiguously determine coupling constants because of the lack of
sufficient experimental data, and therefore a plausible assumption of the SU(3) symmetry
has to be called for. In addition, there is no conclusive way to fix the parameters of the
two-pion exchange $\Lambda NN$ 3BFs that are basically of the NNLO. This means
that experimental and theoretical investigations are needed in the future. It is
essentially important to quantitatively establish the contribution of $\Lambda NN$
3BFs in hypernuclei, which is also relevant for the understanding of the appearance
or absence of $\Lambda$ hyperons in neutron star matter \cite{MK18,GKW20}.

In particular, the investigation of the hypertriton is of fundamental importance.
Before doing full Faddeev
calculations for the hypertriton including $\Lambda NN$ 3BFs,
it is worthwhile to estimate the possible role of the 3BFs for the hypertriton.
The present folding potential calculation indicates the necessity of considering
$\Lambda NN$ 3BFs in the theoretical study of the hypertriton because the
quantitative estimation of their effects will influence the parametrization of
$\Lambda N$ two-body interactions.

\smallskip
{\it Acknowledgements.}
This work is supported by JSPS KAKENHI Grants No. JP19K03849 and No. JP22K03597.

\begin{widetext}
\appendix
\section{Tensor-product  decomposition of $\Lambda NN$ three-body interactions}
 $V_{TPE}^{(K,\ell_a,\ell_b)}(p,q)$ in Eq. (\ref{eq:tpe}) for $K=0$, 1 and 2 in the case of the
leftmost diagram of Fig. 1 are as follows:
\begin{align}
 V_{TPE}^{K=0,\ell_a,\ell_b}(p,q)=&  (-1)^{\ell_a} \delta_{\ell_a\ell_b}\delta_{\ell_a, even}
 \sqrt{\frac{\hat{\ell_a}}{3}} \frac{\{C_0 m_\pi^2+C_1 (-p^2+r_{NN}^2q^2)\}(p^2-r_{NN}^2q^2)}
{2pr_{NN}q(p^2+r_{NN}^2q^2+m_\pi^2)} Q_{\ell_a}(z_{pq}), \\
 V_{TPE}^{K=1,\ell_a,\ell_b}(p,q)=& \delta_{\ell_a\ell_b}\delta_{\ell_a, odd}
 \frac{ \{C_0 m_\pi^2+C_1 (-p^2+r_{NN}^2 q^2)\}}{(p^2+r_{NN}^2q^2+m_\pi^2)}
 \frac{1}{\hat{\ell_a}}\sqrt{\frac{\ell_a(\ell_a+1)}{6\hat{\ell_a}}}
 \{ Q_{\ell_a+1}(z_{pq})- Q_{\ell_a-1}(z_{pq})\}, \\
 V_{TPE}^{K=2,\ell_a,\ell_b}(p,q)=&  -\sqrt{\frac{2}{15}}
 \frac{ \{C_0 m_\pi^2+C_1 (-p^2+r_{NN}^2q^2)\}}{2pr_{NN}q(p^2+r_{NN}^2q^2+m_\pi^2)}
   \sqrt{\hat{\ell_a}\hat{\ell_b}}  (\ell_a 0\ell_b0|20)
  \{ p^2Q_{\ell_b}(z_{pq})- r_{NN}^2  q^2Q_{\ell_a}(z_{pq})\},
\end{align}
where $C_0=-\frac{g_A^2}{3f_0^4}(3b_0+b_D)$, $C_1=\frac{g_A^2}{3f_0^4}(2b_2+3b_4)$, and
$z_{pq}=\frac{p^2+r_{NN}^2q^2+m_\pi^2}{2pr_{NN}q}$. $Q_\ell$ is a Legendre polynomial
of the second kind.

\section{Evaluation of $\Lambda$-deuteron folding potential from $\Lambda NN$ 3BFs}
An explicit calculational procedure of the $\Lambda$-deuteron folding potential,
given by Eq. (\ref{eq:lpd}), is provided. 
The abbreviated notation of Eq. (\ref{eq:lpd}) means
\begin{align}
 U_{\Lambda-d}^{J_t}(q_1',q_1)=& \iint p_1'^2 dp_1' p_1^2 dp_1
 \langle [\Psi_d(\bp_1'),(\ell_{\Lambda}' 1/2)j_\Lambda]J_t|V_{TPE}^{\Lambda NN}
 | [\Psi_d(\bp_1),(\ell_{\Lambda} 1/2)j_\Lambda]J_t  \rangle \notag \\
   =& \sum_{\ell_d'=0,2}\sum_{\ell_d=0,2}\iint p_1'^2 dp_1' p_1^2 dp_1\frac{1}{p_1'}\phi_{\ell_d}(p_1')
   \frac{1}{p_1}\phi_{\ell_d}(p_1)    \langle [ [Y_{\ell_d'}(\hat{\bp_1'}) \times \chi_d^{1}]^1\times
     [Y_{\ell_{\Lambda}'}(\hat{\bq_1'})\times \chi_\Lambda^{1/2}]^{j_\Lambda}]_{M_t}^{J_t}| \notag \\
 & \times     V_{TPE}^{\Lambda NN}  |[ [Y_{\ell_d}(\hat{\bp_1}) \times \chi_d^{1}]^1\times
     [Y_{\ell_{\Lambda}}(\hat{\bq_1})\times \chi_\Lambda^{1/2}]^{j_\Lambda}]_{M_t}^{J_t} \rangle,
\end{align}
where $\chi_d^1$ and $\chi_\Lambda^{1/2}$ denote spin functions for the deuteron and the
$\Lambda$ hyperon, respectively. The isospin degrees of freedom are not explicitly shown.
Because the isospin of the $\Lambda$ hyperon is 0 and that of the deuteron is 0, the matrix
element of the isospin operator $(\bftau_2\cdot\bftau_3)$ in Eq. (\ref{eq:tpe}) becomes $-3$.
Substituting $V_{TPE}^{\Lambda NN}$ of Eq. (\ref{eq:tpe}), the following angular-momentum
recoupling is carried out:
\begin{align}
 & \langle[[Y_{\ell_d'}(\hat{\bp_1'})\times \chi_{d}^1]^1\times
 [Y_{\ell_{\Lambda}'}(\hat{\bq_1'})\times \chi_\Lambda^{1/2}]^{j_\Lambda}]_{M_t}^{J_t}
|V_{TPE}^{(K,\ell_a,\ell_b)}(p,q)[[Y_{\ell_a}(\hat{\bp})\times Y_{\ell_b}(\hat{\bq})]^K
 \times [\bfsigma_2\times \bfsigma_3]^K]_0^0 \notag \\
  & \times|[[Y_{\ell_d}(\hat{\bp_1})\times \chi_{d}^1]^1
 \times [Y_{\ell_{\Lambda}}(\hat{\bq_1})\times\chi_\Lambda^{1/2}]^{j_\Lambda}]_{M_t}^{J_t}\rangle \notag \\
=& \sum_{L'L}\sum_{S'S} 3\hat{j_\Lambda} \sqrt{\hat{L'}\hat{S'}\hat{L}\hat{S}}
 \ninj{\ell_d'}{\ell_{\Lambda'}}{L'}{1}{1/2}{S'}{1}{j_{\Lambda'}}{J_t}
 \ninj{\ell_d}{\ell_{\Lambda}}{L}{1}{1/2}{S}{1}{j_{\Lambda'}}{J_t}
 \langle [[Y_{\ell_d'}(\hat{\bp_1'})\times Y_{\ell_\Lambda'}(\hat{\bq_1'})]^{L'} \times
 [\chi_{d}^1\times\chi_\Lambda^{1/2}]^{S'}]_{M_t}^{J_t}| \notag \\
 & \times V_{TPE}^{(K,\ell_a,\ell_b)}(p,q) [[Y_{\ell_a}(\hat{\bp})\times Y_{\ell_b}(\hat{\bq})]^K
 \times [\bfsigma_2\times \bfsigma_3]^K]_0^0|
 [[Y_{\ell_d}(\hat{\bp_1})\times Y_{\ell_\Lambda}(\hat{\bq_1})]^{L} \times
 [\chi_{d}^1\times\chi_\Lambda^{1/2}]^{S}]_{M_t}^{J_t} \rangle \notag \\
=& \sum_{L'L}\sum_{S'S} 3\hat{j_\Lambda} \sqrt{\hat{L'}\hat{S'}\hat{L}\hat{S}}
 \ninj{\ell_d'}{\ell_{\Lambda'}}{L'}{1}{1/2}{S'}{1}{j_{\Lambda'}}{J_t}
 \ninj{\ell_d}{\ell_{\Lambda}}{L}{1}{1/2}{S}{1}{j_{\Lambda'}}{J_t}
 \sqrt{\hat{J_t}\hat{L'}\hat{S'}} \ninj{J_t}{J_t}{0}{L'}{L}{K}{S'}{S}{K} \notag \\
 & \times  \langle [Y_{\ell_d'}(\hat{\bp_1'})\times Y_{\ell_\Lambda'}(\hat{\bq_1'})]^{L'} ||
 V_{TPE}^{(K,\ell_a,\ell_b)}(p,q) [Y_{\ell_a}(\hat{\bp})\times Y_{\ell_b}(\hat{\bq})]^K ||
 [Y_{\ell_d}(\hat{\bp_1})\times Y_{\ell_\Lambda}(\hat{\bq_1})]^{L} \rangle_{pwe} \notag \\
 & \times 18\sqrt{\hat{S}\hat{K}}(-1)^{K+3/2+S} \sixj{S'}{S}{K}{1}{1}{1/2}
 \ninj{1}{1}{K}{1/2}{1/2}{1}{1/2}{1/2}{1},
\end{align}
where $\bp=\bp_1'-\bp_1$ and $\bq=\bq_1'-\bq_1$. Then, Eq. (\ref{eq:redm}) is applied
in this expression.
The result does not depend on $M_t$. Numerical results of the case
of $\ell_d'=\ell_d=0$ are presented in Sec. IV.\\
\end{widetext}

\end{document}